\documentclass[11pt]{article}
\usepackage{amsfonts}
\usepackage{amsmath}

\newcommand{\sect}[1]{\setcounter{equation}{0}\section{#1}}

\def\bea{\begin{eqnarray}}
\def\eea{\end{eqnarray}}

\def\>#1{{\mathbf#1}}                 

\def\k{{\kappa}}
\def\ka{{a}}
\def\dd{{\rm d}}

\def\y{{q}}
\def\ji{{q}}
\def\m{{\mu}}

\def\la{\lambda}

\begin{document}

 \ 
 \bigskip

\begin{center}
{\Large{\bf{ Superintegrable anharmonic oscillators}}}
 
{\Large{\bf{on $N$-dimensional curved spaces}}}\footnote{Based on the contribution presented at the 
``17th Conference on  Nonlinear
Evolution Equations and Dynamical Systems" NEEDS 2007.
 L'Ametlla  de Mar, Spain,
June 17-24, 2007.}

\end{center}

\bigskip

\begin{center}
{\'Angel Ballesteros$^a$, Alberto Encisco~$^b$, Francisco J. Herranz~$^c$
and Orlando Ragnisco~$^d$}
\end{center}

\noindent
{$^a$ Depto.~de F\'\i sica, Facultad de Ciencias, Universidad de Burgos,
09001 Burgos, Spain\\ ~~E-mail: angelb@ubu.es\\[10pt]
$^b$ Depto.~de F{\'\i}sica Te\'orica II,   Universidad Complutense,   28040 Madrid,
Spain\\ ~~E-mail: aenciso@fis.ucm.es\\[10pt]
$^c$ Depto.~de F\'\i sica,  Escuela Polit\'ecnica Superior, Universidad de Burgos,
09001 Burgos, Spain \\ ~~E-mail: fjherranz@ubu.es\\[10pt]
$^d$ Dipartimento di Fisica,   Universit\`a di Roma Tre and Instituto Nazionale di
Fisica Nucleare sezione di Roma Tre,  Via Vasca Navale 84,  00146 Roma, Italy  \\
~~E-mail: ragnisco@fis.uniroma3.it}

\begin{abstract}
\noindent The maximal superintegrability of the intrinsic harmonic oscillator potential on
$N$-dimensional spaces with constant curvature is revisited from the point of view of
$sl(2)$-Poisson coalgebra symmetry. It is shown how this algebraic
approach leads to a straightforward definition of a new large family of quasi-maximally superintegrable
perturbations of the intrinsic oscillator on such spaces.
Moreover, the generalization of this construction to those $N$-dimensional spaces with non-constant
curvature that are endowed with $sl(2)$-coalgebra symmetry is presented. As the first examples of the latter class of systems, both
the oscillator potential on an
$N$-dimensional Darboux space
as well as several families of its quasi-maximally superintegrable anharmonic perturbations are explicitly
constructed.

\end{abstract}

%%%%%%%%%%%%%%%%%%%%%%%%%%%%%%%%%%%%%%%%%%%%%%%%%%%%%%%%%%%%%%%%%%%%%%%%%%%%%%%%%%

\sect{Introduction}

The Poisson-coalgebraic ``dynamical" symmetry underlying all the superintegrable
Hamiltonian systems that we shall present in the sequel can be summarized as the
following quite general result~\cite{BHletter,sigmaOrlando}: Let $(\>q,\>p
)=\bigl((q_1,\dots,q_N),(p_1,\dots,p_N)\bigr)$ be $N$ pairs of canonical variables
with respect to the  Poisson bracket 
\begin{equation}
\{f,g\}=\sum_{i=1}^N\left(\frac{\partial f}{\partial q_i}
\frac{\partial g}{\partial p_i} -\frac{\partial g}{\partial q_i} 
\frac{\partial f}{\partial p_i}\right),  
\label{ba}
\end{equation} and let us consider the three functions given by
\begin{equation}
  \>q^2=\sum_{i=1}^N q_i^2 , \quad    \>p^2=
    \sum_{i=1}^N    p_i^2   ,  \quad   \>q\cdot\>p =  \sum_{i=1}^N  q_i\, p_i .
\label{bb}
\end{equation} Then, given {\it any} smooth function ${\cal H}$, the Hamiltonian
\begin{equation} H^{(N)}={\cal H}(\>q^2,\>p^2,\>q\cdot\>p)
\label{theham}
\end{equation} defines an $N$-dimensional ($N$D) classical superintegrable Hamiltonian
$H^{(N)}$ with  
$(2N-3)$ functionally independent integrals of the motion that are explicitly   given
by
\begin{equation}
 C^{(m)}= \sum_{1\leq i<j}^m   ({q_i}{p_j} - {q_j}{p_i})^2   , \qquad 
 C_{(m)}=\!\!\! \sum_{N-m+1\leq i<j}^N   ({q_i}{p_j} - {q_j}{p_i})^2   ,
\label{bd}
\end{equation} where $m=2,\dots, N$ and $C^{(N)}=C_{(N)}$.  Furthermore, the sets of
functions given by $\{H^{(N)},C^{(m)}\}$ and $\{H^{(N)},C_{(m)}\}$ $(m=2,\dots, N)$
define two sets of $N$  integrals in involution.
Proofs, technical details and further generalizations can be found
in~\cite{BHletter,sigmaOrlando} but, 
at this point, some remarks concerning the symmetry and
superintegrability properties of
$H^{(N)}$ are in order. 

\begin{itemize}

\item {\bf Remark 1}. If the three functions (\ref{bb}) are written as $J_-$, $J_+$
and $J_3$, respectively, by computing the Poisson bracket~\eqref{ba} among them we recover the  Lie--Poisson commutation rules of $\frak
{sl}(2,\mathbb R)$:
\begin{equation}
 \{J_3,J_+\}=2 J_+     ,\quad  
\{J_3,J_-\}=-2 J_- ,\quad   
\{J_-,J_+\}=4 J_3  .   
\end{equation}
In other words, the functions~\eqref{bb} define a particular $N$D symplectic realization of 
$\frak {sl}(2,\mathbb R)$. Hence any Hamiltonian $H^{(N)}$ can be thought of as a
smooth function defined on $\frak {sl}(2,\mathbb R)$
\begin{equation} 
H^{(N)}={\cal H}(J_-,J_+,J_3)={\cal H}(\>q^2,\>p^2,\>q\cdot\>p),
\label{be}
\end{equation} 
and all the results here presented can be interpreted in the framework
of $\frak {sl}(2,\mathbb R)$-Poisson dynamics. Properly speaking, $H^{(N)}$ is
defined on a three-dimen\-sional $\frak {sl}(2,\mathbb R)$-subalgebra of the $\frak
{sl}(2,\mathbb R)\otimes\cdots\otimes^{N)}\frak {sl}(2,\mathbb R)$ Poisson algebra. We
stress that integrable systems on the Euclidean space and endowed with the $N$-particle $\frak {sl}(2,\mathbb R)$-symmetry given by the  representation (\ref{bb}),  were already studied in~\cite{Wojciechowskia}.

\item {\bf Remark 2}. The ``universal" integrals of motion (\ref{bd}) are derived
from the Casimir function of the aforementioned $\frak {sl}(2,\mathbb R)$ Poisson
algebra (see~\cite{BHletter}), and are given as sums of the square of certain angular momentum
components. In particular, since the functions
$L_{ij}={q_i}{p_j} - {q_j}{p_i}$ with $i<j$ and $i,j=1,\dots,N$ span an $\frak
{so}(N)$ Lie--Poisson algebra with Poisson brackets
\begin{equation}
\{ L_{ij},L_{ik} \}= L_{jk} ,\quad  \{ L_{ij},L_{jk} \}= -L_{ik} ,\quad 
\{ L_{ik},L_{jk} \}= L_{ij} ,\quad i<j<k,
\end{equation}
 the integrals (\ref{bd}) can be rewritten as 
\begin{equation}
C^{(m)}= \sum_{1\leq i<j}^m L_{ij}^2 
,\qquad C_{(m)}=\!\!\! \sum_{N-m+1\leq i<j}^N L_{ij}^2.
\end{equation}
This, in turn, means
that  the $\frak {sl}(2,\mathbb R)\otimes\cdots\otimes^{N)}\frak {sl}(2,\mathbb R)$
symmetry gives us the right prescription to get the appropriate subset of quadratic
functions of the generators of
$\frak {so}(N)$ that Poisson-commute with the Hamiltonian~\eqref{theham} {\it and}
are in involution. In this respect, note that the algebra $\frak {sl}(2,\mathbb
R)\otimes\cdots\otimes^{N)}\frak {sl}(2,\mathbb R)$ has  only $3N$ generators and
many of them do Poisson-commute, whilst $\frak {so}(N)$ has $N(N-1)/2$ generators with
many non-vanishing Poisson brackets among them. 
 
\item {\bf Remark 3}.  It is well-known that the maximum number of functionally independent (and different
from the Hamiltonian itself) integrals of the motion for an $N$D Hamiltonian is
$(2N-2)$. In the case that all these integrals do exist, the sytem is called {\it maximally superintegrable} (MS). Since
$H^{(N)}$~\eqref{theham} has, by construction, $(2N-3)$ functionally independent
integrals, we shall say that this is a {\it quasi-maximally superintegrable} (QMS)
Hamiltonian. Nevertheless, for some specific choices of the function $\cal H$ it will
be possible to find the remaining integral (which is not provided by the
above symmetry). In that case $\cal H$  will define a MS system.

\item {\bf Remark 4}. The canonical variables $(\>q,\>p )$ have {\it a priori} neither
a given geometrical (physical) meaning, nor restricted (real/complex) values.

\end{itemize}

Therefore, we can conclude that  $H^{(N)}$ (\ref{be}) comprises a large family of QMS
Hamiltonians; each particular system arises for a specific choice of the function
$\cal H$ together with an ``appropriate" geometrical interpretation of the canonical
variables.

%%%%%%%%%%%%%%%%%%%%%%%%%%%%%%%

\subsection{Oscillators on the $N$D Euclidean space}

In order to illustrate  these ideas and also as the starting point for further
developments, let us consider the $N$D isotropic harmonic oscillator with angular
frequency $\omega$. Such a system can easily be   identified within the  family 
(\ref{be}) by simply setting
\begin{equation} {\cal  H}  =   \frac1{2} J_+ + \frac1{2} \omega^2 J_- = \frac1{2} 
  \>p^2+ \frac1{2}\omega^2 \>q^2  ,
\label{bf}
\end{equation} for which $\>q$ are Cartesian coordinates in the $N$D Euclidean space
$\mathbb E^N$.
 This Hamiltonian is not only QMS, but the standard prototype of MS systems. In fact, the ``remaining" constant
of the motion can be taken as any of the $N$  integrals
\begin{equation} {\cal I}_i=p_i^2+   \omega^2  q_i^2, \quad i=1,\dots,N,
\label{bg}
\end{equation} since each   ${\cal I}_i$  is functionally independent with respect to
both the set  (\ref{bd}) and ${\cal  H}$.  The results above summarized   allows
for a straightforward superintegrable
 even-order anharmonic oscillator perturbation given by~\cite{BHletter}:
\begin{equation} {\cal  H}  =   \frac1{2} J_+ +  \frac1{2}\omega^2 J_-
+\sum_{k=1}^\infty\delta_k J_-^{k+1} =\frac1{2} 
 \>p^2+ \frac1{2}\omega^2 \>q^2+\sum_{k=1}^\infty\delta_k \>q^{2(k+1)} ,
 \label{bh}
\end{equation} which is QMS for any choice of the $\delta_k$ parameters, since the Hamiltonian~\eqref{bh} is indeed a function of~\eqref{bb}. We remark
that once a single anharmonic contribution with parameter $\delta_k$ is added to the
first harmonic term the maximal superintegrability is lost, but the resulting system
(for any number of arbitrary $\delta_k$'s) always keeps the $(2N-3)$ integrals of
motion (\ref{bd}). In particular,  the latter are just the integrals of the motion
for the radial  
  Garnier system~\cite{ Garnier,Wojciechowski}, which is recovered by taking $\omega$
and $\delta_1$ as the only non-vanishing parameters. We also recall that the integrability properties of some quartic oscillators can be generalized to the Calogero--Moser systems defined with such nonlinear oscillators as external potentals (see~\cite{Wojciechowskib} and references therein).

 In the following sections we present the superintegrable Hamiltonians  defining
anharmonic oscillators on the $N$D  sphere   and hyperbolic  spaces~\cite{Doub}, as
well as on an
$N$D  Riemannian space of variable curvature. The latter space is an $N$D
generalization of the 2D Darboux surface of type III~\cite{Ko72,KKMW03}, one of the
four 2D spaces with non-constant curvature whose geodesic flows are MS. We stress
that in the constant curvature  cases the zero-curvature (flat) limit leads to the Euclidean
nonlinear oscillator (\ref{bh}).

%%%%%%%%%%%%%%%%%%%%%%%%%%%%%%%%%%%%%%%%%%%%%%%%%%%%%%%%%%%%%%%%%%%%%%%%%%%%%%%%%%
\sect{Anharmonic oscillators on spaces of constant curvature}

In this section we consider the $N$D classical Riemannian spaces with 
 constant sectional curvature $\k$: the  sphere $\mathbb S^N$ $(\k>0)$ and the  
hyperbolic space $\mathbb H^N$
$(\k<0)$. We  recall that both of them
 can be embedded in a linear space $\mathbb R^{N+1}$ with ambient or Weierstrass
coordinates $(x_0,\>x)=(x_0,x_1,\dots,x_N)$ subjected to the   ``sphere" constraint
\begin{equation}
\Sigma\ :\quad x_0^2+\k \>x^2=1.
\end{equation}
The metric on the proper $N$D spaces
reads~\cite{VulpiLett}:
\begin{equation}
\dd s^2= {1\over\k}
\left(\dd x_0^2+\k  \dd \>x^2\right)\biggr|_{\Sigma} ,
\label{ca}
\end{equation} where $ \dd \>x^2= \sum_{i=1}^N \dd x_i^2$. 

In order to be able to apply the results described in section 1 to the geodesic flows
and oscillator potentials on $\mathbb S^N$ and $\mathbb H^N$ (with the construction
of the constant curvature counterpart of the QMS Hamiltonian on $\mathbb E^N$
(\ref{bh}) in mind), we shall proceed as follows: 
\begin{itemize}

\item We interpret in a proper way the ``abstract" canonical coordinates and momenta 
$(\>q,\>p )$ as intrinsic quantities on each space; this step can be achieved through 
different projections from the ambient space $\mathbb R^{N+1}$. The resulting metric
in terms of $\>q$ leads to the kinetic energy term of the Hamiltonian.

\item We deduce the form of  the corresponding intrinsic harmonic oscillator potential on the
spaces with constant curvature as a function of
$\>q$; its flat limit (or contraction) $\k\to 0$ has to give (\ref{bf}). Moreover, in our framework 
the well-known maximal superintegrability of this potential on these spaces has to be
explicitly proven by finding the additional integral of the motion through direct computation.  
\item Finally, QMS anharmonic oscillator potentials can be obtained as a symmetry--preserving perturbation of the intrinsic oscillator on these spaces: in particular, we can consider the
sum of all the powers of the above intrinsic oscillator potential in such a way
that  their flat limit $\k\to 0$ reduces to (\ref{bh}). In this way, the constant curvature analogues of the radial Garnier system will be obtained by considering the anharmonicity given by the square of the intrinsic oscillator.

\end{itemize}

In the following we apply the above steps by considering two types of phase
spaces   $(\>q,\>p )$  in  $\mathbb S^N$ and $\mathbb H^N$ coming
from  different  projections from $\mathbb R^{N+1}$. Obviously, the two
Hamiltonians so obtained are canonically equivalent through a change of
coordinates despite its apparent disequivalence as objects defined on the $\frak {sl}(2,\mathbb R)$ Poisson algebra.

%%%%%%%%%%%%%%%%%%%%%%%%%%%%%%%%%%%%%%%%%%%%%%%%%%%%%%%%%%%%%%%%%%%%%%%%%%%%%%%%%%
\subsection{Stereographic projection: Poincar\'e coordinates}

Let us consider the stereographic projection~\cite{Doub}   from the ambient
coordinates
$(x_0,\>x)\in
\Sigma\subset \mathbb R^{N+1}$ to the Poincar\'e coordinates
$\>\y\in \mathbb R^N$ with pole  
$(-1,\>0)\in \mathbb R^{N+1}$:
\begin{equation}
(-1,\>0)+\la\, (1,\>\y)\in\Sigma .
\end{equation}
 Hence we obtain  that
\begin{equation}
\la=\frac{2}{1+  \k\>\y^2}
,\qquad x_0=\la -1=\frac {1-\k\>\y^2}{1+
\k\>\y^2},\qquad 
\>x=\la\,\>\y=\frac{2\>\y}{1+
\k\>\y^2}.
\end{equation}
Therefore, the metric (\ref{ca}) in Poincar\'e   coordinates  reads
\begin{equation}
\dd s^2=4\,\frac{\dd \>\y^2}{(1+\k\>\y^2)^2}  .
\end{equation}
 And the associated geodesic flow has (up to a positive constant
factor) the free Lagrangian 
\begin{equation}
{\cal T}= \frac{\dot{\>\y}^2}{2(1+\k\>\y^2)^2} .
\end{equation}
The    canonical momenta $\>p$ are obtained through the usual Legendre
transformation and read:
\begin{equation}
\>p= \frac{\dot{\>\y}}{(1+\k\>\y^2)^2} .
\end{equation}
Thus the  $N$D kinetic energy  on $\mathbb S^N$ and $\mathbb H^N$ is
given in Hamiltonian form as a particular case of (\ref{be}); namely
\begin{equation}
 {\cal T}=\frac{1}{2}\left( 1+\k J_-\right)^2 J_+=
\frac{1}{2}\left( 1+\k \>q^2\right)^2 \>p^2 .
\label{ce}
\end{equation} 
As a consequence, this geodesic flow is (at least) QMS, with
\eqref{bd} being the explicit (and invariant) form of the integrals of the motion. However, we
stress that in the context of Poincar\'e coordinates the geometric interpretation of such integrals in terms of angular
momentum components is lacking.  

The next point is to deduce the curved harmonic
oscillator potential in terms of these Poincar\'e coordinates $\>q$. We recall  that
the radial (geodesic polar) distance $r$ from an arbitrary point to the origin in
$\mathbb S^N$ ($\k>0$) and   
$\mathbb H^N$ ($\k<0$)  along the geodesic joining both points is written (in  ambient
coordinates) as~\cite{VulpiLett,CRMVulpi} 
\begin{equation}
\frac{1}{\k}\tan^2(\sqrt{\k}\,r) =\frac{\>x^2}{x_0^2}  .
\label{cf}
\end{equation} 
Consequently, the well-known intrinsic oscillator on constant curvature
spaces (the so-called Higgs oscillator~\cite{Higgs,Leemon}) is written in Poincar\'e
coordinates as
\begin{equation} {\cal U}= \frac 12 \omega^2\frac{\>\y^2}{(1-\k\>\y^2)^2} .
\label{cg}
\end{equation} In this way the full Higgs oscillator Hamiltonian reads
\begin{equation}
 {\cal H}=\frac{1}{2}\left( 1+\k J_-\right)^2 J_+ + \frac 12  \omega^2\,  \frac{   J_-}{ (1-\k
J_-)^2 } =  \frac{1}{2}\left( 1+\k \>q^2\right)^2 \>p^2+ \frac 12
\omega^2 \,\frac{\>q^2}{(1-\k\>q^2)^2}  .
\label{ch}
\end{equation} This system is known to be MS~\cite{Higgs}. Therefore, the remaining
functionally independent constant of the motion does exist and, therefore, it has to be found by direct methods. Such an additional integral can be shown to be any
of the following
$N$ functions~\cite{BHletter}:
\begin{equation}
  {\cal I}_i=\left( p_i(1-\k \>q^2) + 2\k (\>q\cdot \>p) q_i 
\right)^2+ \frac{    \omega^2 q_i^2}{(1-\k\>q^2)^2}  ,\quad i=1,\dots,N. 
\label{ci}
\end{equation}

Now, a natural perturbation of this Hamiltonian including anharmonic terms that preserve the QMS properties of the system would be
\bea  &
&{\cal H}=\frac{1}{2}\left( 1+\k J_-\right)^2 J_+ + \frac 12 \omega^2\,  \frac{  
J_-}{ (1-\k J_-)^2 }+\sum_{k=1}^\infty\delta_k \left(  \frac{   J_-}{ (1-\k J_-)^2 }
\right)^{k+1}     
\nonumber\\[2pt] &&\quad =
   \frac{1}{2}\left( 1+\k \>q^2\right)^2 \>p^2+ \frac 12
\omega^2 \,\frac{\>q^2}{(1-\k\>q^2)^2} +\sum_{k=1}^\infty\delta_k \, \frac{ 
\>q^{2(k+1)}}{(1-\k\>q^2)^{2(k+1)}}     
   .
\label{cj}
\eea 
Notice that when any $\delta_k\ne 0$ this curved anharmonic oscillator is QMS
(it always commutes with the integrals~\eqref{bd}  due to its $\frak {sl}(2,\mathbb
R)$-coalgebra symmetry) but not MS (at least with integrals depending quadratically
on the momenta). In fact, the $N=2$ restriction of this perturbed system does not appear in the  classifications 
of MS systems on ${\mathbb S}^2$ and ${\mathbb H}^2$  given
in~\cite{Santandera,Kalninsa,Kalninsb}. Note also that the first perturbative term given by  $\delta_1\ne 0$ can be considered as the constant curvature generalization of the (radial) Garnier system.

%%%%%%%%%%%%%%%%%%%%%%%%%%%%%%%%%%%%%%%%%%%%%%%%%%%%%%%%%%%%%%%%%%%%%%%%%%%%%%%%%%
\subsection{Central projection:  Beltrami coordinates}

Now we consider   the central projection   from the ambient coordinates $ (x_0,\>x)\in
\Sigma\subset \mathbb R^{N+1}$
to the Beltrami ones $\>\ji\in \mathbb R^N$  with pole  
$(0,\>0)\in \mathbb R^{N+1}$: 
\begin{equation}
(0,\>0)+\m\,
(1,\>\ji)\in\Sigma,
\end{equation}
 so that we find  
\begin{equation}
\m=\frac{1}{\sqrt{1+ \k \>\ji^2}},\qquad
 x_0=\m ,\qquad 
\>x=\m\, \>\ji=\frac{\>\ji}{\sqrt{1+ \k \>\ji^2}}.
\end{equation}
Then the metric (\ref{ca})   turns out to be
\begin{equation}
\dd s^2= \frac{(1+\k\>\ji^2)\dd\>\ji^2-\k
(\>\ji\cdot \dd\>\ji)^2}{(1+\k\>\ji^2)^2} .
\end{equation}
The corresponding free Lagrangian is given by
\begin{equation}
{\cal T}=  \frac{(1+\k\>\ji^2)\dot{\>\ji}^2-\k({\>\ji} \cdot \dot{\>\ji})^2}{2(1+\k\>\ji^2)^2},
\end{equation}
which leads to the definition of the conjugate momenta $\>p$ as:
\begin{equation}
  \>p=    \frac{(1+\k\>\ji^2)\dot{\>\ji}-\k (\>\ji\cdot
\dot{\>\ji})\>\ji}{(1+\k\>\ji^2)^2}   .
\end{equation}
Hence the kinetic energy Hamiltonian describing geodesic motion reads
\begin{equation}
 {\cal T}=\frac{1}{2}\left( 1+\k J_-\right)\left(  J_+ +\k
J_3^2\right)=
\frac{1}{2}(1+\k \>q^2)\left( \>p^2+\k (\>q\cdot \>p)^2 \right).
\label{ea}
\end{equation}
By taking into account (\ref{cf}) we find that the  
expression of the curved oscillator on $\mathbb S^N$ and   
$\mathbb H^N$ adopts in these Beltrami coordinates the following ``Euclidean" form:
\begin{equation}
 {\cal U}=  \frac 12\omega^2\>q^2,
\end{equation}
which yields the following expression for the complete curved oscillator Hamiltonian (again as a particular
case of (\ref{be})):
\begin{equation}
{\cal H} =\frac{1}{2}\left( 1+\k J_-\right)\left(  J_+ +\k
J_3^2\right) + \frac 12 \omega^2 J_- =
\frac{1}{2}(1+\k \>q^2)\left( \>p^2+\k (\>q\cdot \>p)^2 \right)+  \frac 12\omega^2 
\>q^2 .
\label{eb}
\end{equation}
The  remaining constant of the motion for this MS Hamiltonian can be taken from any
of the $N$ functions~\cite{BHletter}
\begin{equation}
 {\cal I}_i =\left( p_i+\k (\>q\cdot \>p) q_i \right)^2+
\omega^2  q_i^2,\qquad i=1,\dots,N. 
\label{ec}
\end{equation}
And the explicit QMS anharmonic generalization of (\ref{eb}) is proposed to be
\bea
 &&
{\cal H} =\frac{1}{2}\left( 1+\k J_-\right)\left(  J_+ +\k
J_3^2\right)  + \frac 12 \omega^2 J_-  +\sum_{k=1}^\infty\delta_k J_-^{k+1} \nonumber \\[2pt]
&&\quad 
=
\frac{1}{2}(1+\k \>q^2)\left( \>p^2+\k (\>q\cdot \>p)^2 \right)+ \frac 12 \omega^2 
\>q^2 
+\sum_{k=1}^\infty\delta_k \>q^{2(k+1)} .
 \label{ed}
\eea
Note that in this coordinates the curved Garnier term is given just by the $\>q^4$ perturbation.

%%%%%%%%%%%%%%%%%%%%%%%%%%%%%%%%%%%%%%%%%%%%%%%%%%%%%%%%%%%%%%%%%%%%%%%%%%%%%%%%%%
\sect{Oscillators on an $N$D space of non-constant curvature}

In arbitrary manifolds with non-constant curvature, kinetic-energy Hamiltonians can
exhibit extremely complicated dynamics and are, in general, no longer
integrable~\cite{Pa99}. From a physical viewpoint, the caracterization of
(super)integrable geodesic flows on curved (pseudo)-Riemannian manifolds in arbitrary
dimensions is relevant for supergravity and superstring theories, and constitutes an
active research field (see~\cite{Gibbons} and references therein).   Let us consider
the (spherically symmetric and conformally flat) $N$D Riemannian manifold  whose
metric and geodesic Hamiltonian flow are given by
\begin{equation}
\label{ds2}
\dd s^2=(\ka+\>q^2)\,\dd\>q^2
,\qquad  {\cal T}=\frac{\>p^2}{\ka+\>q^2},
\end{equation} 
where the parameter $\ka>0$.  This is a space with non-constant curvature. Moreover,
its scalar curvature is negative and given by
\begin{equation}
R=-(N-1)\frac{3(N-2)\,\>q^2+2\ka N}{(\ka+\,\>q^2)^3}\,.
\end{equation}
In the $N=2$ case, this space is just one of the so-called 2D Darboux spaces: the
2-manifolds with non-constant curvature admitting two quadratic first integrals, so
that its geodesic motion is quadratically MS. There are only four types of such
spaces~\cite{Ko72}, which are (from the integrability viewpoint) the closest ones
to constant curvature ones, since their $N$D generalizations are the only spaces
other than $\mathbb E^N$, $\mathbb H^N$ and $\mathbb S^N$ whose geodesic motion could
be expected to be (quadratically) MS for any dimension. In fact, an $N$D spherically
symmetric generalization of the four 2D Darboux spaces has been recently introduced
in~\cite{PLB} and shown to be   QMS.

The integrability properties of the space~\eqref{ds2} have been thoroughly studied
in~\cite{PhysicaD},  where it has been shown that the Hamiltonian
\begin{equation}
\label{cH} 
{\cal H}=\frac{J_+
+\omega^2\,J_-}{\ka+J_-}=\frac{\>p^2}{\ka+\>q^2}+
\omega^2\,\frac{\>q^2}{\ka+\>q^2} ,
\end{equation} 
is a MS system with $(2\,N-2)$ functionally independent quadratic
first integrals given again by~\eqref{bd} and  one of the following functions
\begin{equation}
\label{hod3} 
{\cal I}_i =p_i^2- ({\cal H}-\omega^2)\,q_i^2\,, \quad i=1,\dots,N.
\end{equation} 
Note that this integral {\it cannot} be
written as a function of the $\frak {sl}(2,\mathbb R)$ symmetry; moreover, the set
$\{{\cal I}_i:1\leq i\leq N\}$ is also in involution. To the best of our knowledge,
this Hamiltonian provides the first example of a Hamiltonian system on a Riemannian
space of non-constant curvature which is MS in any dimension. 

Furthermore, a geometric analysis shows~\cite{PhysicaD} that the potential 
\begin{equation}\label{V}
{\cal U}=\omega^2\,\frac{J_-}{\ka+J_-}=\omega^2\,\frac{\>q^2}{\ka+\>q^2} ,
\end{equation} 
can be interpreted as the intrinsic harmonic oscillator on this curved
space, that turns out to be MS, despite of the introduction of a non-constant
curvature. We remark that, as expected, for $N=2$ this model is listed in the
classification of MS potentials for the Darboux space of type III given
in~\cite{KKMW03}.   At this point, two different proposals for the definition of an
$N$D QMS anharmonic oscillator perturbation on this space arise in a natural way. The first one
consists in the same type of generalization proposed in the constant curvature cases
described in section 2:
\bea  
&&{\cal H}=\frac{J_+}{\ka+J_-} + \omega^2\,\frac{J_-}{\ka+J_-}
+\sum_{k=1}^\infty\delta_k \left(  \frac{J_-}{\ka+J_-}\right)^{k+1}     
\nonumber\\[2pt] &&\qquad =
\frac{\>p^2}{\ka+\>q^2}+
\omega^2\,\frac{\>q^2}{\ka+\>q^2}\,
 +\sum_{k=1}^\infty\delta_k \left( \frac{\>q^2}{\ka+\>q^2}  \right)^{k+1}     
   .
\label{and3a}
\eea A second (and perhaps more natural) possibility arises if we realize that
in~\eqref{cH} the intrinsic oscillator on this curved space is just the ``Euclidean"
oscillator divided by the conformal factor of the metric, which carries all the
information concerning the non-constant curvature of the space  (see~\cite{PLB}).
From this perspective, the curved anharmonic oscillator would be defined as
\begin{equation}
\label{and3b} 
{\cal H}=\frac{J_+ + \omega^2\,J_-
+\sum_{k=1}^\infty\delta_k   {J_-}^{k+1}}{\ka+J_-} =\frac{\>p^2 +
\omega^2\,\>q^2 +\sum_{k=1}^\infty\delta_k  
{\>q^{2(k+1)} } }{\ka+\>q^2}.
\end{equation} 
In any case, both Hamiltonians~\eqref{and3a} and~\eqref{and3b} are QMS
and, like the rest of the systems presented in this paper, they do have the same set
of universal integrals~\eqref{bd} coming from their $\frak {sl}(2,\mathbb R)$
symmetry.  

%%%%%%%%%%%%%%%%%%%%%%%%%%%%%%%%%%%%%%%%%%%%%%%%%%%%%%%%%%%%%%%%%%%%%%%%%%%%%%%%%%
%%%%%%%%%%%%%%%%% Acknowledgments %%%%%%%%%%%%%%%%%%

\section*{Acknowledgments}

\noindent
 This work was partially supported by the Spanish MEC and by
the Junta de Castilla y Le\'on under grants no.\ FIS2004-07913, MTM2007-67389 and
VA013C05 (A.B.\ and F.J.H.), by the Spanish DGI under grant no.\ FIS2005-00752 (A.E.)
and by the INFN--CICyT (O.R.).  A.E. acknowledges the financial support of the
Spanish MEC through an FPU scholarship. A.B.\ and F.J.H.\ are also grateful to G.S.
Pogosyan for helpful discussions.

%%%%%%%%%%%%%%%%%%%%%%%%%%%%%%%%%%%%%%%%%%%%%%%%%%%%%%%%%%%%%%%%%%%%%%%%%%%%%%%%%%

\label{lastpage}

\end{document}